# HARMONIC ANALYSIS OF TIME VARIATIONS OBSERVED IN THE SOLAR RADIO FLUX MEASURED AT 810 MHZ IN THE YEARS 1957-2004.

S. Zięba, J. Masłowski, A. Michalec, G. Michałek, AND A. Kułak
Astronomical Observatory, Jagiellonian University, ul. Orla 171, 30-244 Kraków, Poland

ABSTRACT

Long-running measurements of solar radio flux density at 810 MHz were analyzed in order to find similarities and differences among solar cycles and possible causes responsible for them. Detailed frequency, amplitude and phase characteristics of each analyzed time series were obtained by using modified periodograms, an iterative technique of fitting and subtracting sinusoids in the time domain, and basing on the least squares method (LSM). Solar cycles 20, 21, 22 and shorter segments around solar minima and maxima were examined separately. Also, dynamic studies with 405, 810 and 1620 days windows were undertaken. The obtained harmonic representation of all these time series indicates large differences among solar cycles and their segments. These findings show that the solar radio flux at 810 MHz violate the Gnevyshev Ohl rule for the pair of cycles 22-23. Analyzing the entire period 1957-2004 the following spectral periods longer than 1350 day: 10.6, 8.0, 28.0, 5.3, 55.0, 3.9, 6.0, 4.4 14.6 yr were detected. For spectral periods between 270 and 1350 days the 11 years cycle is not recognized. We think that these harmonics form “impulses of activity” or quasi-biennial cycle defined in the Benevolenskaya model of the “double magnetic cycle” (Hale’s 22 yr cycle and quasi-biennale cycle). A value of about 0.09 is proposed for the interaction parameter between the low and high-frequency components of the Benevolenskaya model. Our findings confirm the intermittent behavior of the periodicity up to nearly 155 days. Correlation coefficients between daily values of the radio emission at 810 MHz and sunspot numbers as well as the radio emission at 2800 MHz calculated for 540 days interval depend on the solar cycle phase. The largest values (~0.9) are for the rising and declining phases of sunspot cycle but significantly smaller (~0.5) for intervals around the minimal and maximal phases of cycle.

## 1. INTRODUCTION

The origin and persistency of solar activity is one of the basic problems of solar physics. Generally it is assumed, that the local and global fields which are responsible for various manifestations (symptoms, signs) of the solar activity, are generated by the solar magnetic cycle. It consists of two about 11-year cycles of sunspot activity and affects not only all levels of the Sun but also the entire heliosphere including the Earth. The 22 yr cycles begin on even sunspot cycles, according to the Zürich numbering, and manifest themselves in a reversal of the polarity of sunspots (Hale’s law). This reversal corresponds to the change of the toroidal magnetic field. The large-scale (background) magnetic field regions of the Sun also show a 22-year periodicity. It also changes polarity but only at the maximum of sunspot activity (Makarov et al. 2001).

Such a periodic behavior of the sunspot and background fields generally agrees with the Babcock-Leighton semi-empirical theory (Leighton, 1969). The weak, large-scale, high-latitude magnetic field on the Sun is explained as the result of the accumulation of field transported poleward from lower latitude after a breakup of strong fields of active regions (Leighton 1964; Wang, Nash & Sheeley 1989; Wang & Sheeley 1991; Wang, Lean & Sheeley 2002). However, many questions, e.g., concerning the duration of solar cycles, the reason for impulses of activity (quasi-biennial cycle) and other observed periodicities, the existence of a complex structure of the maximum phase of the sunspot cycle, the single or triple polar magnetic field reversals, remain open.

The activity of the Sun is expressed through various solar indices, which reflect the direct observed solar features (e.g. sunspots, plages, flares) or measure solar emissions at different wavelengths (e.g. the 2800 MHz radio flux, the 530.3 nm, Fe XIV coronal line intensities, the total solar irradiance, the X-ray 0.1-0.8 nm flux). Many of them have been established very recently and unfortunately their short time behavior is generalized for a long period of solar activity, which is inappropriate in the nonstationary process developed on the Sun. One of the best-known and longest activity indices, which has exact physical meaning in contrast to the Wolf number, is the 10.7 cm (2800 MHz) solar radio flux measured daily by the National Research Council of Canada since 1947, February 14.

In the absence of radio bursts, the total solar flux measured in the wavelength range of 1-100 cm can be decomposed into a slowly-varying component SVC, which changes over hours to years, and quiet-Sun radiation $Q_0$ corresponding to the emission from the Sun at the minimum of activity. The SVC was first discovered by Covington (1947, 1948) at 10.7 cm, and by Lehaney & Yabsley (1949) at 25 and 50 cm. At different wavelengths in the 1-100 cm range, various emission mechanisms are responsible for contributions to the SVC. At the 10.7 cm wavelength, the SVC may comprise a varying mix of contributions: thermal free-free emission, thermal gyroresonance emission, and possibly non-thermal gyrosynchrotron emission (Tapping, Cameron & Willis 2003). This mixture of possible emission mechanisms makes studies of the long-term processes of solar activity based on the 10.7 cm data difficult to decipher. However, at wavelengths longer than 20 cm the emission is almost completely due to thermal free-free emission (Nitta et al., 1991; Gopalswamy, White & Kundu, 1991), which makes all interpretations easier.

In our solar radio observations at 37 cm (810 MHz), we investigate time variations of magnetic structures developing in the solar atmosphere. This is possible because increases in radio emission (in the absence of radio bursts) at 37 cm can be explained by thermal free-free emission from concentrations of plasma supported by magnetic loops. Using the classical least squares method, the modified periodograms and an iterative technique of fitting and subtracting sinusoids in the time domain (Delache & Scherrer 1983; Horne & Baliunas 1986; Crane 2001), we analyzed solar activity variations in various phases of sunspot cycle and in different time scales. This allows us to find differences between solar sunspot cycles (cycles 20-22) and distinguish their various multi-peaked temporal structures with small amplitudes not connected with the 11-year sunspot cycle. As our observations cover almost 50 years, we suppose that the analysis of different time series created from the data can help in the understanding the solar activity variations.

## 2. OBSERVATIONS

Regular, everyday observations of solar radio emissions at the Astronomical Observatory of Jagiellonian University at 810 MHz began on 1957, October 1. For this purpose, the parabolic 7-meter antenna was used in an equatorial mounting and the total power radiometer (Czyżewski 1958, Machalski & Urbanik 1975). On 1 March 1996, we started new accurate measurements of the total solar radio flux at ten frequencies within the decimeter range of wavelengths (one of them being 810 MHz) using the 8-meter radio telescope built in 1995 at the Krakow Astronomical Observatory (Zięba et al. 1996). Recently, from all the observations carried out from 1957, October 1 until 2004, June 30, a consistent time series was compiled (Zięba et al. 2005) (http:/www.oa.uj.edu.pl). The obtained daily values of solar radio flux at 810 MHz are plotted in Figure 1a. The data cover the period from the maximum of the 19th solar cycle till the declining phase of the 23rd cycle (altogether 17075 days). From extrapolation to zero activity, we estimate the quiet-Sun level $Q_0$ equal to 35 solar flux units. It is observed at the minima of activity where there are no flares and the number of active regio is minimal. In such a circumstance, free-free emission from thermal plasma is the operative mechanism. The active regions emission in excess (in the absence of radio bursts) comprises two main contributions: emission from sources in active regions and any extended emission from outside of active regions. Both the contributions can be accounted for in terms of thermal, free-free emission from concentrations of plasma supported by magnetic loops associated with sunspots and other magnetic structures.

Panels b, c, d of Figure 1 show the correlation coefficient between the daily radio flux at 810 MHz and the international solar numbers. The coefficient was calculated for 270 day (10 solar rotations) intervals, sliding every 10 days along the data (Figure 1b). The coefficient fluctuates between values 0.4 - 0.85. However, the time intervals, when these extremely values are noticed, are not a randomly distributed but appear in definite phases of the sunspot cycle. The maximal values become visible in the rising or declining phases of sunspot cycle, when systematic long-period (longer then ~270 days) increases or decreases of activity dominate over short-period (smaller then ~100 days) variations. During the time of maxima and minima the correlation coefficients are lower and their values depend on how large are contributions to the radio emission from concentration of plasma located in magnetic loops associated with sunspots. Results of some additional calculations of the correlation coefficient between these data are presented in Figures 1c and 1d. Figure 1c gives the coefficients calculated when the radio data was shifted with respect to sunspot numbers. The mutual 10-day shifts of these data cause large differences in the run of the correlation coefficient. The evident decrease of the correlation indicates that most of the observed radio emission at 810 MHz is generated in the active regions. Figure 1d presents the correlation coefficient computed for 540 days intervals. These values are also plotted in Figure 2, but the phase of the solar cycle stands in for the time. The phase locations of the computed values of the correlation coefficient support our conclusion that their values do not change accidentally with the phase of the sunspot cycle. All the cycles have the largest coefficients near the phases 0.1-0.25 and 0.6-0.75. However, for some cycles large values of the coefficient are also noticed beyond these intervals which indicated differences among cycles. This effect is seen also when the correlation coefficient is calculated for longer intervals (810 days).

## 3. HARMONIC ANALYSIS OF THE DATA

Solar activity is not a stationary or temporally homogeneous process, so the behavior of various solar indices depends upon the time interval under study. Therefore, analysis of solar data sets applies only locally and for comparative studies it is important to have a proper method of describing observed data. It is difficult to find differences or similarities among time series created from various sections of the data in the time domain, so for this purpose a spectral analysis is rather applied. As the three techniques (correlation, power-spectrum and modified periodogram) generally used in spectral analysis have some disadvantages (Crane 2001), in our approach we applied the classical least squares method for finding sinusoids, which together with a linear function reconstructs an investigated time series.

### 3.1 *Decomposition of time series into harmonics*

Although our approach is based on the classical least squares method (LSM), the modified periodograms and an iterative technique of fitting and subtracting sinusoids in the time domain (Horne & Baliunas 1986) are used to establish initial frequencies and descending amplitude order of successive harmonics.

We start our considerations by calculating the Scargle normalized periodogram $P_N(\omega)$ (Scargle 1982). The highest peak in the original data periodogram provides the frequency $\omega_1$ corresponding to the strongest sinusoidal signal present in the data. Then, using LSM the phase and amplitude of this sinusoidal signal are fitted from the original data. The sine curve found in such a way is subtracted from the data and a new recalculated periodogram specifies a frequency $\omega_2$ of a subsequent harmonic. In the next step, LSM is again used for the original data in order to find the best fitted phases and amplitudes of the two harmonics with frequencies $\omega_1$ and $\omega_2$. This procedure is repeated so long as the variance of residuals (the difference between the original data and these computed from a set of found harmonics and a linear trend) was equal to about 1% of the original data variance. In some cases all the significant sinusoids hidden in the data were found before this limit was obtained. In the last step of our calculation LSM is used not only for recalculation of phases and amplitudes but also to calculate the frequencies, up to now fixed on the values obtained from periodograms. After this procedure, we find all the harmonic parameters, which are the best fit to the original data. We used the similar approach in our paper analyzing periodicities in solar data during the minimum and the rising phase of solar cycle 23 (Zięba et al. 2001).

### 3.2 *Clusterization of harmonics*

Because there are at least a few dozen detected harmonics numbers in the series to be discussed below, we gathered them into six spectral groups (bands) according to their periods. We recognize the harmonics with periods longer than 1350 days (about 50 solar rotations) as the first group (called L group), which describes the intensity of the low-frequency component (Hale's 22 yr cycle) connected to the dynamo action located near the bottom of the convection zone (Benevolenskaya 1998, 2003). The harmonics from the second group (called I) have periods longer than 270 days (~10 solar rotations) and shorter than 1350 days. We associate them with impulses of activity or a quasi-biennial cycle responsible for the double-peak structure of the 11 yr sunspot cycle. The third group (called A group) comprises of harmonics with periods between 35 and 270 days, while the fourth group (R group) consists of periods inside the 25–35 day time interval of the Sun's rotation. The harmonics with shorter periods belong to the fifth and sixth group. They are periods from 18 to 25 days (S group), and from 7 to 18 days (M group). In respect of a correlation between the solar daily data measured on days not too far one from another, we did not search for harmonics with periods shorter then 7 days (Oliver & Ballester 1995).

The sum of the squared amplitudes of all the harmonics from a given group (the group sum called SSA) can be used as a measure of the activity strength carried by this group. Adding all the SSA together we get a measure of the total activity strength of the analyzed time series (SSAT). As the SSAT values for all the series are dominated by amplitudes of the long periods harmonics, we also looked into the sum of the squared amplitudes but calculated from the harmonics which belong only to the five spectral groups from the band I to M (SSAI-M).

### 3.3 *Determination of investigated time series*

When non-stationary processes are investigated, each analyzed time series must be precisely defined. This is done by establishing the beginning and end of a chosen data set. Below we define 12 different time series, all constructed from daily values of solar radio flux at 810 MHz. The longest one covers all the data. The next three series are comprised the data from the sunspot cycles 20, 21 and 22 respectively (marked as series 20, 21, 22). Their beginnings and ends are defined by the minima determined from the plot drawn on the basis of the linear trend and all the sinusoids having periods longer then 1350 days (white line in Fig.1a) which were found in the all data set. We choose this procedure, because the solar cycles so defined are clearly determined and unbiased by shorter then 1350 day fluctuations of the activity. The successive eight series are described as time intervals between the days in which the sunspot number was equal to 0 the first time before the minimum of activity and the last time after it. A time interval is called passive if it includes a sunspot minimum or active when it contains a maximum. These series are marked by the sunspot cycle number and the letter 'p' or 'a'. The letter indicates what kind of series, passive or active, is considered. For example 20p marks the passive period around the minimum between the solar cycles 19 and 20. Beginnings and durations of the analyzed time series are given in Table 1.

### 3.4 *Harmonic representation of time series*

The harmonic representation of all the investigated time series allows us to study similarities and differences among them much easier than in the case when they would be analyzed directly. Table 1 accumulates information about parameters pertaining to the harmonic representation of the 12 time series. As calculations of harmonics were undertaken so long as the ratios of the residual variances to the original ones had the similar values for all the series ( the ratio values are inside the interval 0.9% - 1.7%) we can assume the similar accuracy of our considerations. In consequence, if a few time series with similar sizes are characterized by different numbers of harmonics, then the series with the largest number of harmonics has the most complicated time structure. It results from the fact, that decays or amplifications of a given periodicity along an investigated time series must be realized by a group of harmonics having very different phases but almost equal frequencies and similar amplitudes. However, it does not mean that the strength of such a series (SSAT) is also the largest. This can be seen when we compare, for example, the solar cycles 20 and 22. According to the data presented in the Table 1 and 2 the 20th cycle is described by 214 harmonics with the total strength 628. This value is about 2.3 times smaller than SSAT of the cycle 22 described by 95 harmonics. Having the harmonic representation of an analyzed time series it is possible to present how large are contributions to the series from the above defined groups of harmonics. This is shown in Figure 3 constructed from the 520 harmonics hidden in the total set of the data. Figures 4 and 5 present the similar harmonic representation but for passive and active segments of the data respectively, computed from the found frequencies, amplitudes and phases, separately for each of the defined group of periodicities. The differences among all these segments are pronounced in the spectral groups M, S and R, but this is not so clear in the case of harmonics with periods between 35 and 270 days.

TABLE 1

PARAMETERS CHARACTERIZING THE 12 ANALYZED TIME SERIES AND THEIR HARMONIC REPRESENTATION

| NAME OF SERIES | START OF SERIES | SIZE OF SERIES [DAYS] | CORRELA TION COEFICIENT WITH ISN | OTT | MEAN VALUE OF SERIES [S.F.U.] | ORIGINAL VARIANCE OF SERIES $[S.F.U.]^2$ | RATIO OF VARIANCES (RESIDUAL TO ORIGINAL) [%] | NUMBER OF DETECTED HARMONICS | AMPLITUDE OF THE SMALLEST HARMONIC [S.F.U.] | VALUE OF LINEAR TREND IN MIDLLE OF SERIES [S.F.U.] |
|---|---|---|---|---|---|---|---|---|---|---|
| **T** | 1 | 17075 | 0.91 | 0.96 | 66.7 | 583.0 | 1.2 | 520 | 0.17 | 66.3 |
| **20** | 2551 | 4170 | 0.88 | 0.94 | 59.4 | 267.2 | 1.0 | 214 | 0.18 | 59.2 |
| **21** | 6721 | 3690 | 0.91 | 0.96 | 68.1 | 584.3 | 1.0 | 118 | 0.34 | 68.6 |
| **22** | 10411 | 3570 | 0.92 | 0.97 | 66.9 | 684.5 | 1.0 | 95 | 0.39 | 70.0 |
| **20p** | 1493 | 1744 | 0.71 | 0.88 | 42.5 | 34.2 | 1.7 | 112 | 0.10 | 43.5 |
| **21p** | 5775 | 1457 | 0.72 | 0.85 | 41.5 | 24.7 | 1.7 | 105 | 0.10 | 41.9 |
| **22p** | 9549 | 1331 | 0.84 | 0.93 | 42.5 | 64.4 | 0.9 | 80 | 0.14 | 42.2 |
| **23p** | 13333 | 1379 | 0.66 | 0.87 | 41.4 | 20.9 | 1.8 | 93 | 0.10 | 42.8 |
| **20a** | 3237 | 2538 | 0.76 | 0.88 | 68.8 | 154.3 | 1.4 | 209 | 0.15 | 81.1 |
| **21a** | 7232 | 2317 | 0.83 | 0.91 | 82..6 | 310.6 | 1.0 | 146 | 0.24 | 80.3 |
| **22a** | 10880 | 2453 | 0.86 | 0.94 | 78.9 | 523.6 | 1.0 | 102 | 0.35 | 78.1 |
| **23a** | 14712 | 2208 | 0.75 | 0.90 | 79.3 | 264.8 | 1.1 | 136 | 0.21 | 78.0 |

The name of series stand for following investigated time series: **T**- the total set of the data, **20** – the solar cycle 20, **21** – the solar cycle 21, **22** – the solar cycle 22, **20p** – a time interval around the minimum between the solar cycles 19 and 20 (from the first to the last day without sunspot), **21p**, **22p**, **23p**, the same as 20p but around the minima between the solar cycles 20 and 21, 21 and 22, 22 and 23 respectively, **20a** – a time interval around the maximum of the solar cycle 20, **21a**, **22a**, **23a** the same as 20a but around the maxima of the solar cycles 21, 22 and 23, respectively.

ISN – the International Sunspot Number. OTT – the solar radio flux at 2800 MHz (10.7 cm).

For the correlation coefficient 0.66 (the worst case - the series **23p**) the lower and upper limits of the 95% probability confidence interval are 0.66-0.03 and 0.66+0.03 respectively.

TABLE 2

NUMBERS OF HARMONICS AND SUMS OF THEIR SQUARED AMPLITUDES FOR SIX SELECTED SPECTRAL GROUPS AND FOR ALL THE INVESTIGATED TIME SERIES

| SERIES | **L** | | **I** | | **A** | | **R** | | **S** | | **M** | | **L-M** | **I-M** |
|---|---|---|---|---|---|---|---|---|---|---|---|---|---|---|
| | NR | SSA | NR | SSA | NR | SSA | NR | SSA | NR | SSA | NR | SSA | SSA | SSA |
| **T** | 9 | 948.4 | 35 | 81.3 | 206 | 45.0 | 119 | 51.0 | 86 | 9.1 | 65 | 3.6 | 1138 | 190 |
| **20** | 3 | 476.7 | 8 | 45.9 | 60 | 35.8 | 32 | 47.3 | 39 | 13.5 | 72 | 9.1 | 628 | 152 |
| **21** | 2 | 1002.9 | 6 | 61.9 | 51 | 74.7 | 22 | 59.8 | 20 | 10.6 | 17 | 5.2 | 1215 | 212 |
| **22** | 2 | 1226.4 | 6 | 77.7 | 37 | 35.7 | 24 | 85.8 | 15 | 9.3 | 11 | 2.9 | 1438 | 211 |
| **20p** | 1 | 63.6 | 4 | 17.7 | 25 | 10.1 | 13 | 14.4 | 17 | 3.3 | 52 | 2.3 | 111 | 48 |
| **21p** | 1 | 31.3 | 3 | 11.5 | 24 | 6.4 | 13 | 15.8 | 15 | 5.6 | 49 | 2.1 | 73 | 41 |
| **22p** | 1* | 30.8 | 3 | 31.0 | 23 | 8.0 | 10 | 7.2 | 13 | 2.1 | 30 | 2.0 | 81 | 50 |
| **23p** | 1 | 34.4 | 3 | 3.3 | 23 | 5.8 | 12 | 6.9 | 14 | 1.6 | 40 | 1.4 | 53 | 19 |
| **20a** | 1 | 90.6 | 6 | 61.0 | 46 | 36.9 | 22 | 63.3 | 30 | 16.7 | 104 | 15.3 | 284 | 193 |
| **21a** | 1 | 442.7 | 6 | 130.4 | 38 | 92.4 | 19 | 72.7 | 20 | 15.4 | 62 | 14.0 | 767 | 324 |
| **22a** | 1 | 1024.0 | 5 | 136.3 | 33 | 49.4 | 19 | 115.0 | 19 | 14.8 | 25 | 8.2 | 1348 | 324 |
| **23a** | 1 | 356.1 | 5 | 124.9 | 33 | 61.0 | 18 | 77.6 | 19 | 18.1 | 60 | 12.5 | 650 | 294 |

We use the following abbreviation for the sixth selected groups of periods (periods are measured in days): **L** – long periods (P=>1350), **I** – impulsive periods (270<=P<1350), **A** – active periods (35<=P<270), **R** – rotational periods (25<=P<35), **S** – short periods (18<=P<25) and **M** – magnetic periods (7<=P<18).

NR – the number of harmonics inside a given group.

SSA – the sum of the squared amplitudes of all the harmonics having periods within the selected period group (measured in $[s.f.u.]^2$).

* the period of the denoted harmonic (P=1345) is a bit out of the group limit, but it is accepted as a long period.

TABLE 3

THE PERIODS AND AMPLITUDES OF THE FIVE STRONGEST HARMONICS IN EACH OF THE SPECTRAL GROUP FOR ALL THE ANALYZED TIME SERIES

| PERIODS SERIES Δf [nHz] | **L** >1350 DAYS | [S.F.U.] | **I** 270÷1350 DAYS | [S.F.U.] | **A** 35 ÷ 270 DAYS | [S.F.U.] | **R** 25 ÷ 35 DAYS | [S.F.U.] | **S** 18 ÷ 25 DAYS | [S.F.U.] | **M** 7 ÷ 18 DAYS | [S.F.U.] |
|---|---|---|---|---|---|---|---|---|---|---|---|---|
| **T** | 3888 | 28.1 a | 979 | 3.4 a | 199 | 1.6 a | 28.0 | 1.5 a | 24.93 | 0.8 a | 13.8 | 0.4 a |
| 0.23 | 2935 | 8.3 a | 1161 | 3.0 a | 198 | 1.3 a | 27.4 | 1.4 a | 24.88 | 0.7 a | 16.8 | 0.4 a |
| | 10242 | 6.7 a | 545 | 2.5 a | 260 | 1.3 a | 27.0 | 1.4 a | 23.7 | 0.6 a | 13.66 | 0.3 a |
| | 1939 | 4.8 a | 950 | 2.4 a | 230 | 1.3 a | 26.7 | 1.2 a | 24.8 | 0.6 a | 13.2 | 0.3 a |
| | 20094 | 4.0 a | 556 | 2.4 a | 225 | 1.2 a | 27.2 | 1.2 a | 23.9 | 0.5 a | 13.71 | 0.3 a |
| **20** | 3953 | 21.1 a | 656 | 3.7 a | 209 | 1.7 a | 28.0 | 2.5 a | 24.9 | 1.9 a | 13.8 | 0.8 a |
| 0.92 | 2024 | 5.1 a | 967 | 3.7 a | 262 | 1.5 a | 27.5 | 2.5 a | 24.7 | 1.0 a | 14.3 | 0.7 c |
| | 1393 | 2.7 a | 534 | 3.0 a | 125 | 1.4 a | 25.9 | 2.0 a | 23.9 | 1.0 a | 16.9 | 0.6 d |
| | | | 280 | 2.2 a | 180 | 1.3 a | 26.8 | 2.0 a | 20.7 | 0.8 a | 16.7 | 0.6 d |
| | | | 456 | 1.3 a | 165 | 1.3 a | 27.2 | 2.0 a | 24.6 | 0.7 a | 13.7 | 0.6 d |
| **21** | 3673 | 31.4 a | 933 | 4.7 a | 267 | 3.2 a | 27.9 | 3.9 a | 24.9 | 1.3 a | 13.7 | 1.0 b |
| 1.05 | 1458 | 3.9 a | 325 | 4.4 a | 156 | 2.7 a | 27.0 | 3.1 a | 24.5 | 1.1 a | 13.9 | 0.8 b |
| | | | 449 | 2.5 a | 187 | 2.5 a | 26.8 | 2.4 a | 23.6 | 1.1 a | 13.4 | 0.7 b |
| | | | 484 | 2.4 a | 224 | 2.5 a | 28.2 | 2.2 a | 22.8 | 1.0 a | 13.3 | 0.7 b |
| | | | 376 | 2.3 a | 242 | 2.5 a | 25.2 | 1.8 a | 24.3 | 0.9 b | 15.0 | 0.6 c |
| **22** | 3200 | 34.8 a | 946 | 6.8 a | 259 | 2.3 a | 26.2 | 3.3 a | 24.4 | 1.2 a | 13.8 | 0.8 a |
| 1.08 | 1449 | 3.7 a | 393 | 3.3 a | 192 | 1.6 a | 26.5 | 3.2 a | 23.1 | 1.2 a | 13.6 | 0.6 c |
| | | | 695 | 3.2 a | 159 | 1.4 a | 29.1 | 2.8 a | 23.8 | 1.0 a | 17.6 | 0.6 c |
| | | | 346 | 2.5 a | 178 | 1.4 a | 25.6 | 2.7 a | 23.6 | 1.0 a | 18.0 | 0.5 d |
| | | | 291 | 1.6 a | 219 | 1.3 a | 28.8 | 2.6 a | 24.1 | 0.8 a | 15.2 | 0.5 d |
| **20p** | 1993 | 8.0 a | 561 | 3.6 a | 110 | 1.3 a | 28.9 | 1.8 a | 21.4 | 0.9 a | 17.8 | 0.4 d |
| 2.21 | | | 901 | 1.8 a | 90 | 1.2 a | 29.3 | 1.8 a | 21.2 | 0.9 a | 15.7 | 0.4 d |
| | | | 322 | 1.3 a | 116 | 1.0 a | 27.5 | 1.4 a | 23.8 | 0.4 d | 14.6 | 0.4 e |
| | | | 273 | 0.2 e | 186 | 0.9 a | 28.0 | 1.2 a | 20.1 | 0.4 e | 17.3 | 0.4 e |
| | | | | | 150 | 0.8 a | 32.2 | 1.0 a | 18.4 | 0.4 e | 15.2 | 0.3 e |
| **21p** | 1524 | 5.6 a | 672 | 2.4 a | 37 | 0.9 a | 31.1 | 2.3 a | 24.99 | 1.5 a | 13.8 | 0.4 d |
| 2.65 | | | 372 | 2.1 a | 254 | 0.8 b | 31.5 | 2.0 a | 23.2 | 1.0 a | 13.9 | 0.4 d |
| | | | 509 | 1.1 a | 71 | 0.8 c | 25.4 | 1.9 a | 24.0 | 0.7 c | 15.7 | 0.4 d |
| | | | | | 125 | 0.8 c | 32.1 | 1.1 a | 20.2 | 0.7 c | 17.6 | 0.3 e |
| | | | | | 52 | 0.7 d | 26.9 | 0.7 c | 22.1 | 0.6 c | 14.5 | 0.3 e |
| **22p** | 1345* | 5.6 a | 550 | 3.8 a | 232 | 1.3 a | 26.3 | 1.5 a | 24.0 | 0.8 a | 13.2 | 0.6 d |
| 2.90 | | | 401 | 3.7 a | 90 | 1.0 a | 27.3 | 1.5 a | 21.6 | 0.5 b | 13.3 | 0.5 d |
| | | | 324 | 1.8 a | 106 | 0.9 a | 27.7 | 0.9 a | 18.1 | 0.4 d | 14.6 | 0.3 e |
| | | | | | 172 | 0.9 a | 32.7 | 0.7 a | 18.3 | 0.4 d | 15.6 | 0.3 e |
| | | | | | 140 | 0.8 a | 29.4 | 0.6 a | 23.2 | 0.4 e | 13.0 | 0.3 e |
| **23p** | 1766 | 5.9 a | 480 | 1.7 a | 48 | 0.8 a | 25.8 | 1.2 a | 19.9 | 0.5 b | 13.8 | 0.4 c |
| 2.80 | | | 709 | 0.7 c | 94 | 0.8 a | 27.6 | 1.2 a | 24.5 | 0.5 b | 11.2 | 0.3 e |
| | | | 340 | 0.3 e | 230 | 0.7 a | 28.0 | 0.9 a | 23.5 | 0.5 c | 15.9 | 0.3 e |
| | | | | | 262 | 0.7 a | 26.9 | 0.8 a | 22.3 | 0.4 d | 13.2 | 0.3 e |
| | | | | | 133 | 0.7 a | 29.3 | 0.8 a | 18.4 | 0.4 e | 16.9 | 0.3 e |
| **20a** | 2453 | 9.5 a | 635 | 4.2 a | 207 | 2.2 a | 27.4 | 3.9 a | 24.9 | 2.4 a | 13.8 | 1.2 b |
| 1.52 | | | 949 | 3.9 a | 83 | 1.7 a | 27.9 | 3.5 a | 20.6 | 1.2 b | 13.7 | 1.0 d |
| | | | 525 | 3.6 a | 180 | 1.6 a | 26.1 | 2.9 a | 23.9 | 1.2 b | 16.9 | 1.0 d |
| | | | 273 | 3.5 a | 99 | 1.6 a | 26.8 | 2.2 a | 24.3 | 0.9 d | 16.7 | 0.9 d |
| | | | 347 | 1.8 a | 41 | 1.4 a | 28.6 | 2.0 a | 19.6 | 0.9 d | 13.2 | 0.9 d |
| **21a** | 2545 | 21.0 a | 1181 | 8.1 a | 231 | 3.9 a | 28.0 | 4.6 a | 24.4 | 1.8 a | 13.7 | 1.1 b |
| 1.66 | | | 341 | 5.4 a | 188 | 3.6 a | 27.0 | 3.8 a | 23.5 | 1.6 a | 13.9 | 1.0 b |
| | | | 648 | 4.7 a | 261 | 3.5 a | 25.7 | 2.4 a | 22.7 | 1.3 a | 13.3 | 0.9 b |
| | | | 595 | 3.4 a | 164 | 2.3 a | 25.1 | 2.2 a | 24.8 | 1.2 b | 15.0 | 0.9 c |
| | | | 293 | 1.0 b | 153 | 2.3 a | 27.6 | 2.2 a | 20.5 | 1.0 b | 13.8 | 0.8 c |
| **22a** | 2479 | 32.0 a | 1211 | 8.4 a | 266 | 3.3 a | 26.1 | 4.3 a | 24.4 | 1.5 a | 13.8 | 1.1 a |
| 1.57 | | | 722 | 5.6 a | 185 | 2.2 a | 26.6 | 4.3 a | 23.2 | 1.4 a | 13.6 | 1.0 b |
| | | | 362 | 5.1 a | 153 | 1.8 a | 29.0 | 4.0 a | 23.7 | 1.3 a | 17.6 | 0.8 b |
| | | | 335 | 2.3 a | 103 | 1.5 a | 25.7 | 3.7 a | 22.9 | 1.2 a | 15.1 | 0.7 c |
| | | | 446 | 1.9 a | 81 | 1.4 a | 28.2 | 2.7 a | 24.0 | 1.2 a | 13.2 | 0.6 d |
| **23a** | 2404 | 18.9 a | 1058 | 6.8 a | 144 | 2.9 a | 27.4 | 4.0 a | 20.9 | 1.9 a | 12.5 | 0.9 c |
| 1.75 | | | 600 | 6.4 a | 112 | 2.8 a | 26.7 | 3.6 a | 23.9 | 1.8 a | 14.2 | 0.8 d |
| | | | 387 | 4.5 a | 165 | 2.6 a | 25.5 | 2.7 a | 24.8 | 1.3 a | 13.1 | 0.8 d |
| | | | 335 | 3.7 a | 130 | 2.3 a | 27.5 | 2.7 a | 21.9 | 1.3 a | 12.8 | 0.8 d |
| | | | 281 | 2.0 a | 234 | 2.3 a | 31.4 | 2.4 a | 21.6 | 1.1 c | 11.9 | 0.8 e |

Δf-denotes the uncertainty in frequency due to the length of the time series.
The letter a, b, c, d, e standing after the amplitude marks to which of the interval defined below belongs the false alarm probability (FAP) value of the given harmonic computed according to the procedure described by Bai and Cliver (1990). The interval a) is for FAP<0.005%, b) for 0.005%<= FAP <0.05%, c) for 0.05%<= FAP <0.5%, d) for 0.5%<= FAP <5 %, e) for 5%<= FAP <50 %,

The sums of harmonics from the spectral group I are characterized by two or three maxima observed in each of the analyzed segments of the data. However, oscillations seen in the passive segments are smaller than those occurred in the active ones.

It is obvious, that we cannot present parameters of all the found harmonics but we list in the Table 3 periods and amplitudes of the five strongest sinusoids from each of the six spectral groups. We also mark the statistical significance level of each presented harmonic giving its false alarm probability (FAP) that is the probability that the harmonic is generated by noise (Bai and Cliver 1990, Bai 1992). FAP values of some harmonics (mainly those having small amplitudes) are rather large (between 5 and 50 %), so they should not be treated as separate and independent time structure but as components of one complicated structure described by a set of harmonics with similar periods. Such situations appear particularly among harmonics with shortest periods. However, these harmonics should be taken into account if we want to represent a time series with sufficient precisely through a set of harmonics. Figure 6 illustrates this approach for the case of time series 23p. Considering the period P of a single harmonic even with a very small FAP value it is necessary to keep in mind the period uncertainty ΔP due to the length τ of the time series. If we accept the formula Δf = 11574/(3τ) (τ is measured in days and f in nHz) for the frequency uncertainty Δf (de Jager 1987) then the range of period variability can be calculated from the following formulas $P_{min}$ = 11574/(11574/P+Δf) and $P_{max}$ = 11574/(11574/P-Δf) where P is the formally calculated value listed in Table 3. Also Δf value for each of the investigated time series is given below the series name in the first column of Table 3, what allows us to compute the period range. For example, if we take the period 401 days found in the series 22p then $P_{min}$ = 364 and $P_{max}$ = 446 days.

## 4. DYNAMIC STUDIES

Because of the non-stationary nature of the solar activity, we examine also how the short-term, "dynamical" behavior of the data changes from time interval to time interval during the years 1957-2004. For this purpose, harmonic analysis is performed for shorter intervals (smaller subsets) of data using a sliding window. In this way computed sets of harmonics demonstrate a dynamical behavior of the original time series. We perform these studies for the 405, 810 and 1620 days windows changing their positions along the time axis every the time interval equal about 1/9 of the windows width. Figure 7 presents the results of the dynamic calculation for the 1620 days sliding window, while the next two (Figures 8 and 9) show these obtained for the 810 and 405 days windows respectively. The horizontal lines of used grids are drawn in the distances equal to the window dimension. The most prominent feature in the figures 7-9 are wavy structures at many periods running vertically from bottom to top. Often these structures follow one another. This can be clearly noticed near the time of the minima of activity. At periods smaller than about 1 year the structure of the 11 years sunspot cycle is seen in the figures as changing numbers of observed sinusoids. However, for longer periods the 11 years cycle is not recognized in that analysis.

## 5. CONCLUDING DISCUSSION

We studied temporal variations of the solar radio emission considering the Krakow Sun observations at 810 MHz for the period 1957-2004. Solar cycles 20, 21 and 22 as well as segments of the data around solar minima and maxima were examined separately. Also dynamic studies with 405, 810 and 1620 days windows were undertaken. In each case the investigated time series was transformed into its harmonic representation, what allows to study similarities and differences between various segments of the data much easier than in the case when they would be analyzed directly.

The basic question discussed in this work can be summarized as follows:

1) Daily values of the solar radio emission at 810 MHz shown in Figure 1a indicate the typical variability of solar activity. The correlation coefficient between these data and the international solar numbers is equal 0.91 when the whole period 1957-2004 is analyzed. For smaller data segments values of the correlation coefficient fluctuate with a phase of solar cycle. The maximal values appear in the rising or declining phases of sunspot cycle (see Figure 1d and 2), when systematic longer than about 10 solar rotations increases or decreases of activity dominate over short-period (smaller then ~100 days) variations. During the time of maxima and minima the correlation coefficients are lower and their values depend on how large are contributions to the radio emission from concentration of plasma located in magnetic loops associated with sunspots. The correlation coefficients between the daily values of solar radio emission at 810 MHz and 2800 MHz are larger than those calculated between our data and sunspot numbers (see Table 1) but their variations with a phase of solar cycle are similar.

So, we think that quoting correlation coefficients as a result of correlation study between some solar indices of activity both investigated time intervals and solar cycle phases should be defined.

2) The harmonic representations of all the investigated segments of the data show rather large numbers of sinusoids needed for reconstructions of analyzed time series. It is obvious, that these numbers change for different considered time segments. However, if the accuracy of the time series reconstruction (for example measured by the ratio of the reconstructed time series variance to the original one) is the same for different investigated series having similar sizes, then the found numbers of harmonics indicate which series have more or less complicated time structure. It results from the fact, that decays or amplifications of a given periodicity along an investigated time series, must be realized by a group of harmonics having very different phases but almost equal frequencies and similar amplitudes. Results of our calculation, given in Table 1 and 2, clearly state that the simplest time structure of the observed solar radio emission occurs in cycle 22 while the most complicated one appears during cycle 20. Concurrently, the total activity strength of cycle 22, measured by the sum of the squared amplitudes of all the harmonics describing the series (TSSA), is more than 2 times greater than for cycle 20. This ratio is even bigger (~5) when the active segments of the both cycles are compared. The largest contribution comes from the harmonics with periods longer than 1350 days (the L group) but the observed tendency remains similar for shorter spectral groups (see Table 2). Comparing the TSSA values of the four successive active segments around the maxima of solar cycles 20, 21, 22, 23 we see that relation between TSSA in the pair 22a-23a is opposite to that in the pair 20a-21a. This is in agreement to sunspot observations which also show for the pair of cycles 22-23 the violation of the Gnevyshev-Ohl rule (Gnevyshev & Ohl 1948, Duhau 2003). From Figures 3, 4 and 5 it is evident that the differences among the investigated cycles and among passive and active segments of the solar cycles are pronounced mainly in two groups of harmonics: the R group (periods from 25-35 days) and the S group (periods from 18 to 25 days). For cycle 22 the structure generated by the R group harmonics is clearly different from those generated by other cycles. Perhaps, the compact structure given by the harmonics from this group cause cycle 22 to violate the Gnevyshev-Ohl rule.

Table 3 presents the five harmonics with the largest amplitudes from each of the spectral group and for all the investigated segments of the data. Generally amplitudes of the five strongest harmonics from a given spectral group have similar amplitudes. This occur mainly for harmonics from the spectral group M having periods between 7 and 18 days. Coming to the groups with longer periods we observe a larger scatter of amplitudes.

From the harmonic analysis of the entire period 1957-2004 the following spectral periods longer than 1350 days are detected: 10.6 [10.7], 8.0 [7.6], 28.0 [30.3], 5.3 [5.3], 55.0 [60.5], 3.9 [3.8], 6.0, 4.4 14.6 [15.1] yr. In square brackets we quote Fourier periods of the Morlet wavelet components which represent yearly mean time series of geomagnetic index aa and Wolf sunspot number for the time period 1844-2000 (Duhau and Chen 2000, Duhau 2003). Interestingly there is a quite good agreement of periods obtained from the analysis of daily solar radio emission and yearly mean data of geomagnetic index and sunspot number. The bottom panel in Figure 3 presents the sum of 9 the longest harmonics mentioned above superposes on the linear trend found in the data.

Since 1984 the attention has been focused on a periodicity near 154 days discovered by Rieger et al.(1984) in the record of solar flare activity from 1980 to 1983. The similar periodicities (152, 153, 155, 158 days) were found in many different indicators of solar activity (see for example Dennis 1985, Bogart & Bai 1985, Lean & Brueckner 1989, Bai & Cliver 1990, Dröge et al.1990, Pap, Tobiska and Bouver 1990, Carbonell & Ballester 1990,1992, Bai & Sturrock 1991,1993, Kile & Cliver 1991, Bouver 1992, Oliver, Ballester, Baudin 1998, Ballester, Oliver, Baudin 1999). It is not possible in this paper to discuss all the problems connected with the near 154 days periodicity. In our data, the strongest harmonic which can be identified with this periodicity is the harmonic having the amplitude 0.84 s.f.u. and the period 156 days (FAP<0.005%). In total, inside the period interval between 150 and 160 days there are 5 harmonics (151,153,156,158,159 days) all with FAP smaller than 0.01%. Figure 10 shows sums constructed from the above mentioned harmonics. The sum of the harmonics 153 and 156 has the largest oscillations during maximum phases of the solar cycles 19, 21 and 23 and indicates some 22 years time structure. This structure is changing when the remaining three harmonics are added. However, again the largest oscillations are observed in 21 cycle (years 1980-81). It is difficult to qualify harmonics to a set which reconstructs a time structure connected with some physical process. To see what is going when the interval 150-160 days is enlarged, we present in Figure 10 sums of harmonics from intervals of 145-165 and 140-170 days. The largest oscillations occur during the maximum of the 21st solar cycle in all the discussed sets of harmonics but smaller oscillations are also present in other segments of the data. The analysis of periods and amplitudes of the strongest line presented in Table 3 confirms this picture. Concluding, our data shows that the periodicity near 155 days belongs to intermediate-term periodicities which activity is time dependent. The intermittent behavior of this periodicity was stated by Antalová (1999) from Fourier analysis of the non-flare full-disk soft X-ray background and the LDE-type flare occurrence.

Because of the non-stationary nature of the solar activity, we undertake also a study of short-term, "dynamical" behavior of the data during the years 1957-2004. For this purpose, harmonic analysis is performed for shorter subsets of data, selected using a sliding window. We perform these studies for the 405, 810 and 1620 days windows changing their positions along the time axis. Intervals use for a window displacement are chosen as equal about 1/9 of the windows width. A careful analysis of Figures 7-9 allows us to present the following conclusions: 1) The most prominent feature in these figures are wavy structures at many periods running vertically from bottom to top. Individual elements of these structures span one or two windows dimension. Structures connected with short periods (between 8 and 20 days) are less wiggly than those related to periods between 20 and 40 days and spread from minimum to minimum. Often these structures follow one another. In the range of periods between 40 and 100 days, their regular changes going on mainly from longer to shorter periods can be noticed. This happens during all the cycles except the cycle 21 (Figure 7). Analyzing the periodicities near to 155 days, we state differences among pictures obtained from different windows. In the case of the 405 day window a dispersion of periods between ~130 and ~180 days and its patchy structure is clearly visible. This can be caused rather by the uncertainty due to the short time window than by the real changes of harmonic periods. When longer windows are used periods from this range show more regular runs but evidently different in various phases of the solar activity. So, the dynamic study confirms also the intermittent behavior of about 155 day periodicity. 2) At periods smaller than about 1 year the structure of the 11 years sunspot cycle is seen as changing numbers of sinusoids appearing in harmonic representations of solar cycle segments. However, for longer periods the 11 year cycle is not recognized.

The numbers of harmonics necessary to represent any time series with a good accuracy (so called representative harmonics) are numerous. It is rather obvious that not all of them represent real physical processes going on in the solar atmosphere, so the basic problem is how to find out harmonics connected with main physical mechanisms of solar activity. It is a real challenge to all who study the periodicities observed from different indices of solar activity. In our opinion many representative harmonics appear as a result of some modulation processes in which primary carrier frequencies comes from rotation periods of solar atmosphere while modulations signals are from the fluctuating radio emission of radio active region having shorter or longer life time. Our simulation studies show that this is possible and will be present in a paper in preparation.

3). The harmonics with periods between 270 and 1350 days (spectral group I) do not manifest the 11 years cycle

(see Figures 3, 6 and 7) but Figure 4 and 5 show maxima (usually two) occurring both at the sunspot cycle maximum and near it minimum. We think that these harmonics form “impulses of activity” proposed by Gnevyshev (Eigenson et al., 1948, Antalova & Gneyshev 1985, Benevolenskaya 2003). These impulses can be explained by the high-frequency component of the toroidal magnetic field in the double magnetic cycle model (Benovolenskaya 1998). The strength (SSA) of harmonics from group I is small in comparison with those connected with large harmonics (group L). The ratio of SSAI to SSAL equals 0.086 (0.092 if sinusoids with periods>22 yr are eliminated) for the total investigated time period and take values 0.096, 0.062, 0.063 for solar cycles 20, 21 and 22 respectively. We propose this ratio as a measure of the interaction between two independent component of the magnetic field defined in the Benevolenskaya model of the “double magnetic cycle”: a low-frequency component (Hale’s 22 yr cycle) generated by the dynamo action located near the bottom of the convection zone and a high-frequency component (impulses of activity or quasi-biennial cycle) produced by the dynamo operating near the top of the zone.

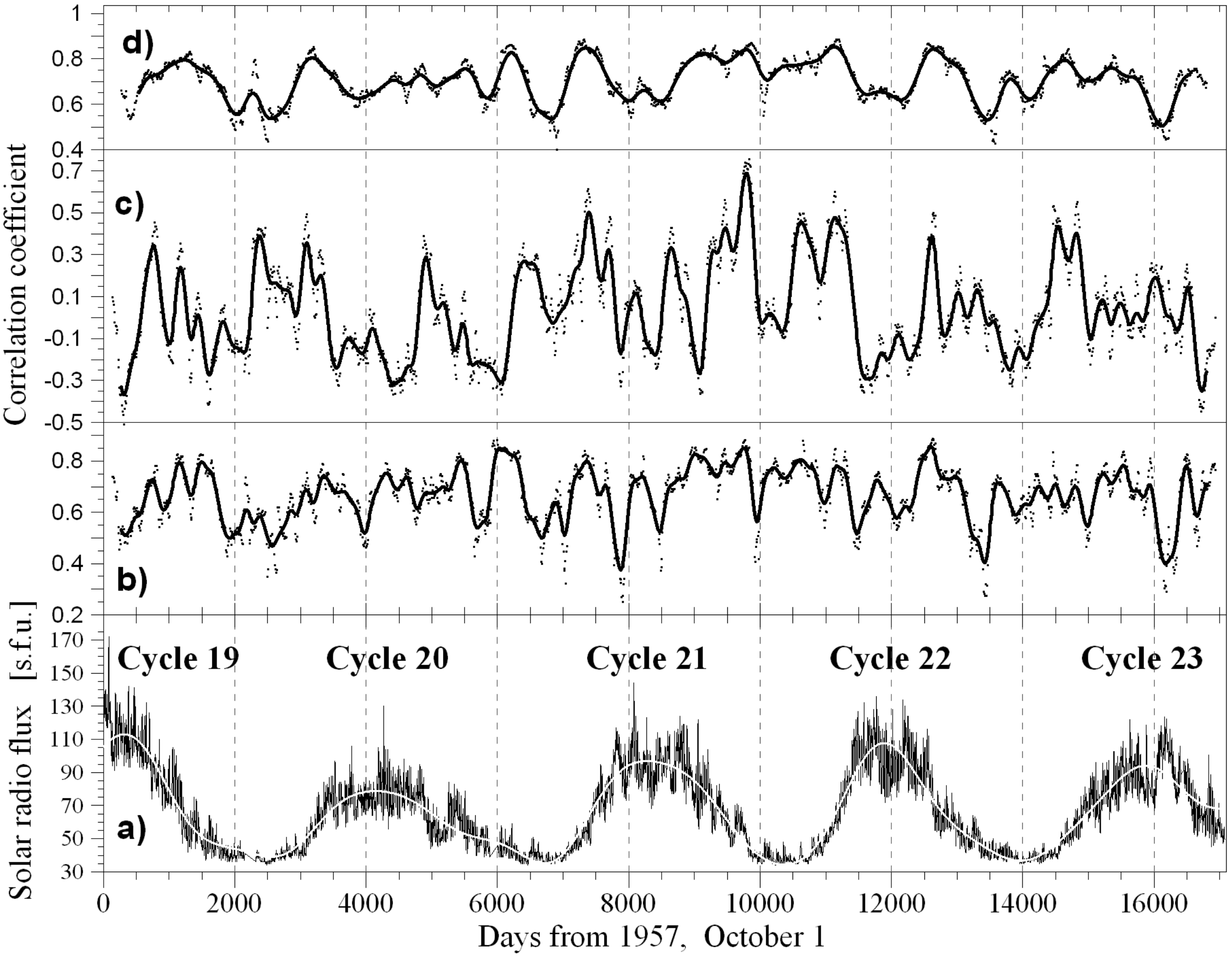

FIG. 1.—a) Plot of the daily values of the total solar flux measured at 810 MHz. The white line shows values calculated from linear trend and all the harmonics, hidden in these data, having periods longer then 1350 days (~50 solar rotations). b) The correlation coefficient between the daily radio flux at 810 MHz and the international solar numbers. Points give the values calculated for 270 days interval (sliding every 10 days) and the line presents 140-day Gaussian average. c) The same as (b) but the radio data are shifted ten days with respect to the sunspot numbers. d) Points give the values calculated for 540 days interval and the line shows 270-day Gaussian average. Letters m and x indicate approximate positions of solar cycles minima and maxima.

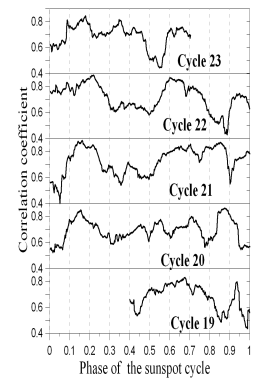

FIG. 2.—The correlation coefficient between the daily radio flux at 810 MHz and the international solar numbers calculated for 540 days intervals plotted against the sunspot cycle phase. Subsequent panels present the correlation coefficient for different solar cycles. For the correlation coefficient 0.8, the lower and upper limits of the 95% probability confidence interval are 0.77 and 0.83 respectively, while for the coefficient 0.5 the limits are 0.44 and 056.

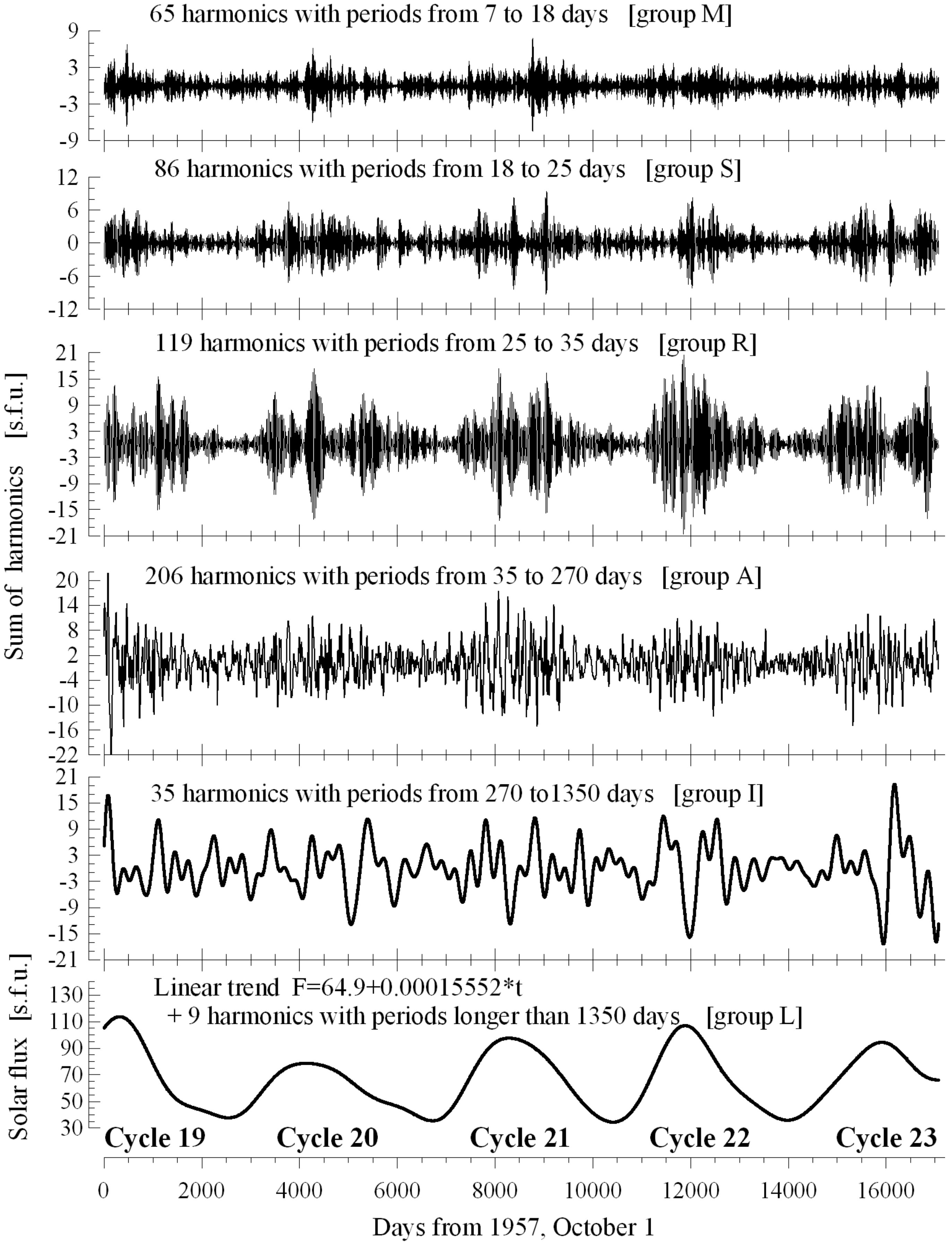

FIG. 3.—Harmonic representation of the time series involving the all observations. Subsequent panels present sums of the harmonics, having periods from the indicated intervals.

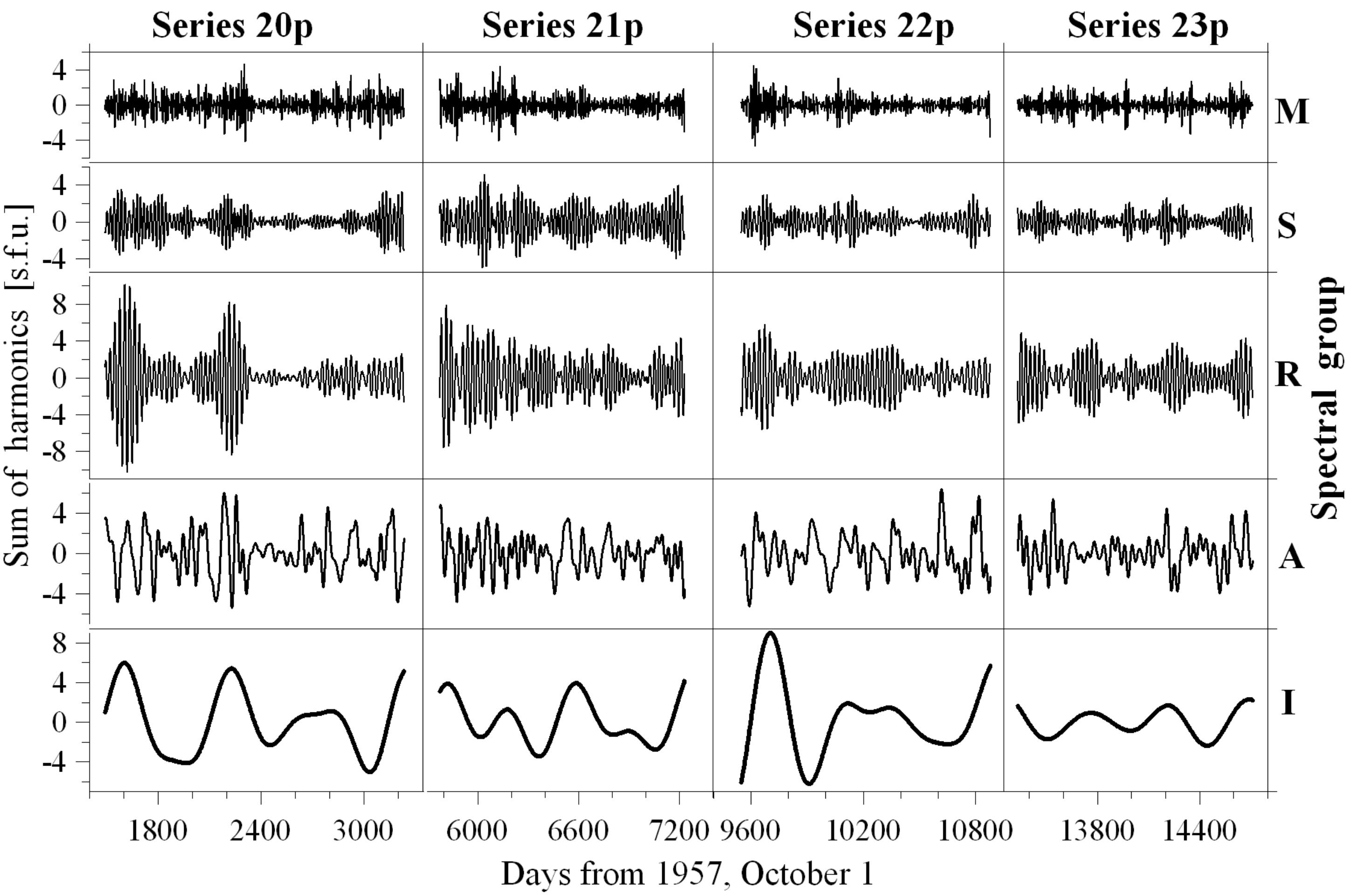

FIG. 4. —Harmonic representations of the four passive segments of the data. Subsequent panels present sums of the harmonics, having periods from the different spectral group, which are marked by the letters **I, A, R, S, M**, where **I** – impulsive periods (270<=P<1350), **A** – active periods (35<=P<270), **R** – rotational periods (25<=P<35), **S** – short periods (18<=P<25) and **M** – magnetic periods (7<=P<18).

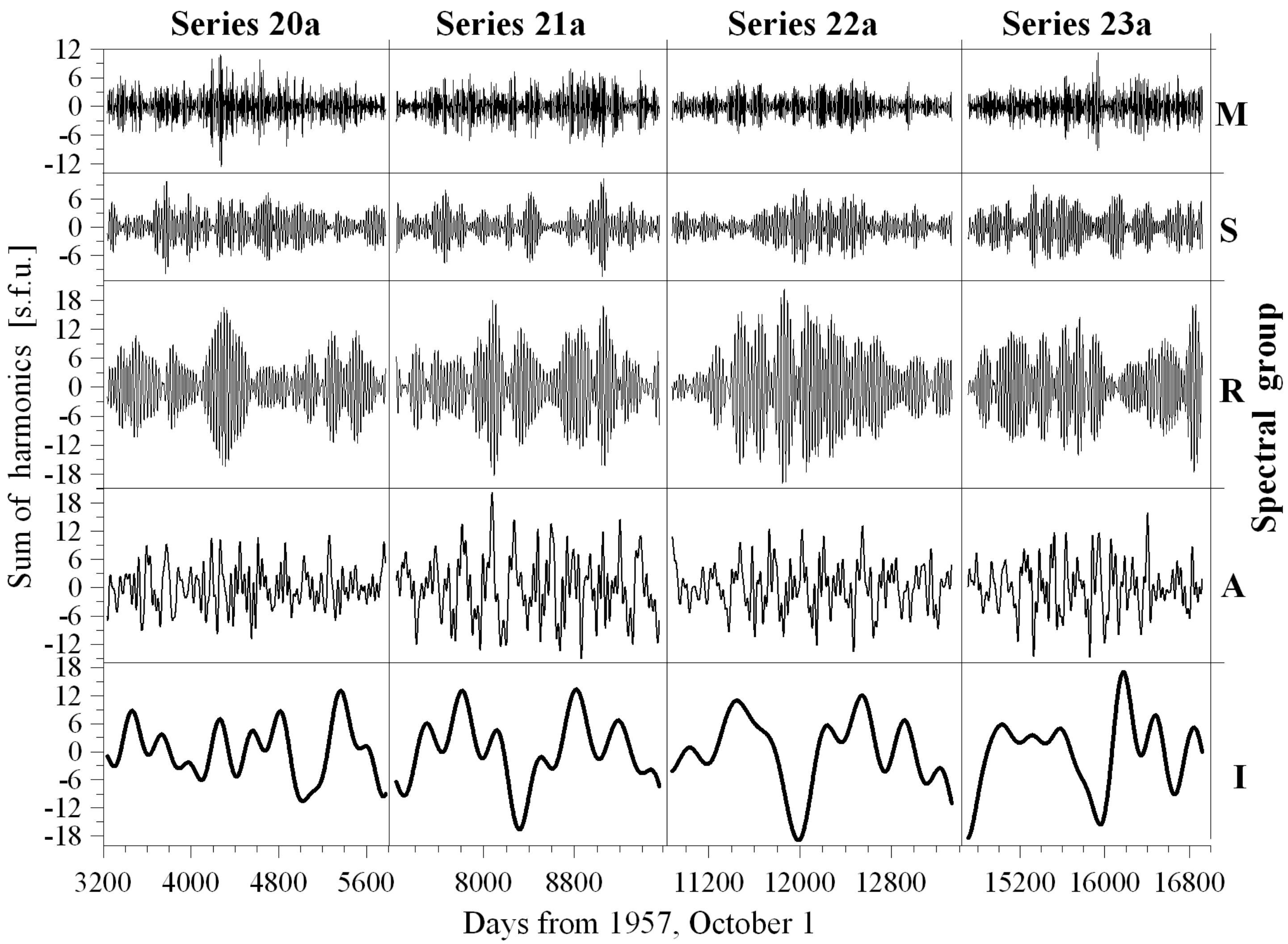

FIG. 5. —The same as Fig.4 but for the active segments of the data.

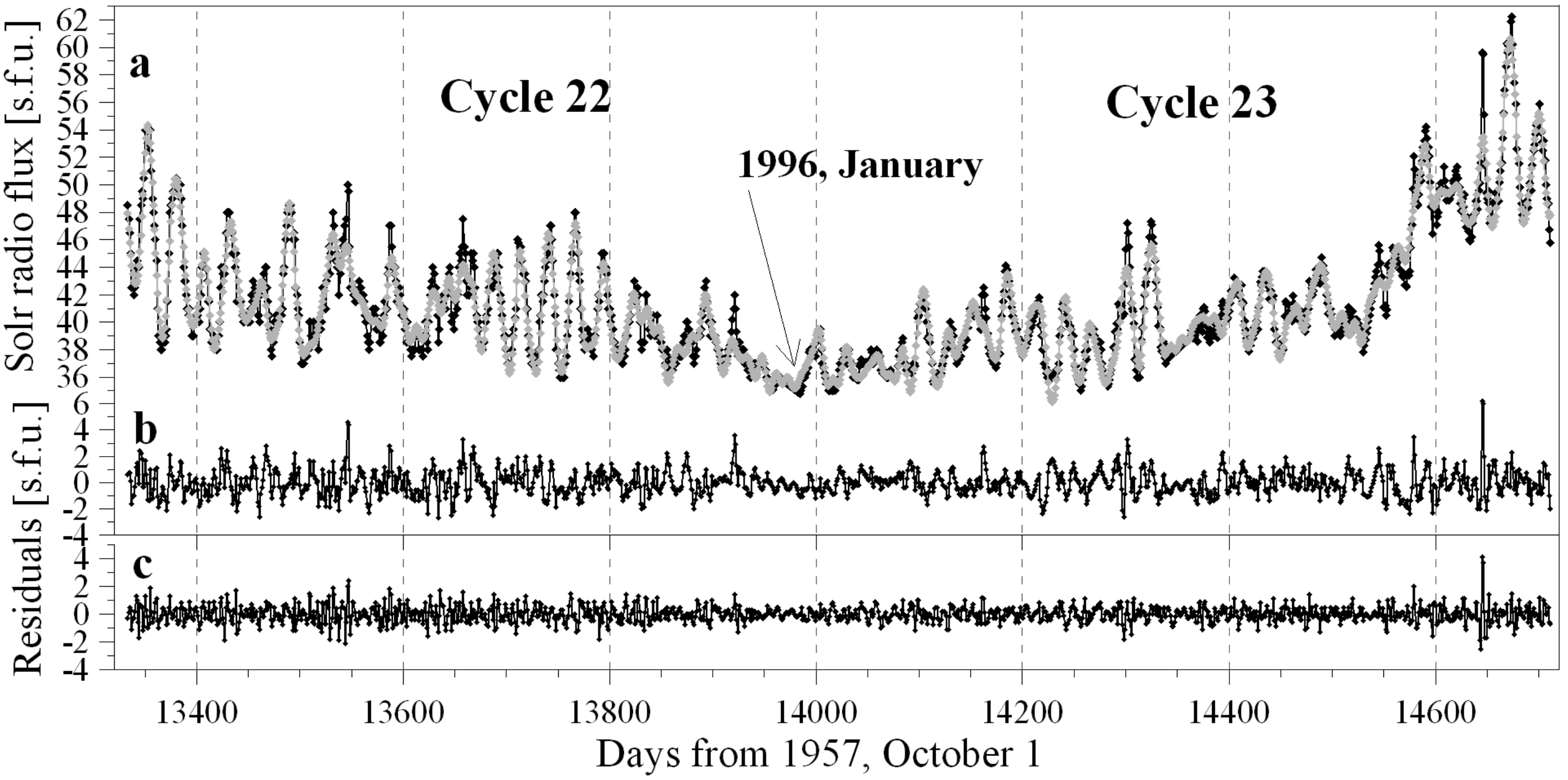

Fig. 6. —a) The observed (black points) and reconstructed from the linear trend and 46 harmonics with FAP smaller than 5% (gray points) daily values of the total solar flux at 810 MHz drawn for the time interval between solar cycle 19 and 20 (the time series 23p). b) Residuals resulting from differences between the observed and reconstructed daily values presented in (a). c) The same as (b), but the reconstructed daily values of solar flux were calculated from 93 harmonics.

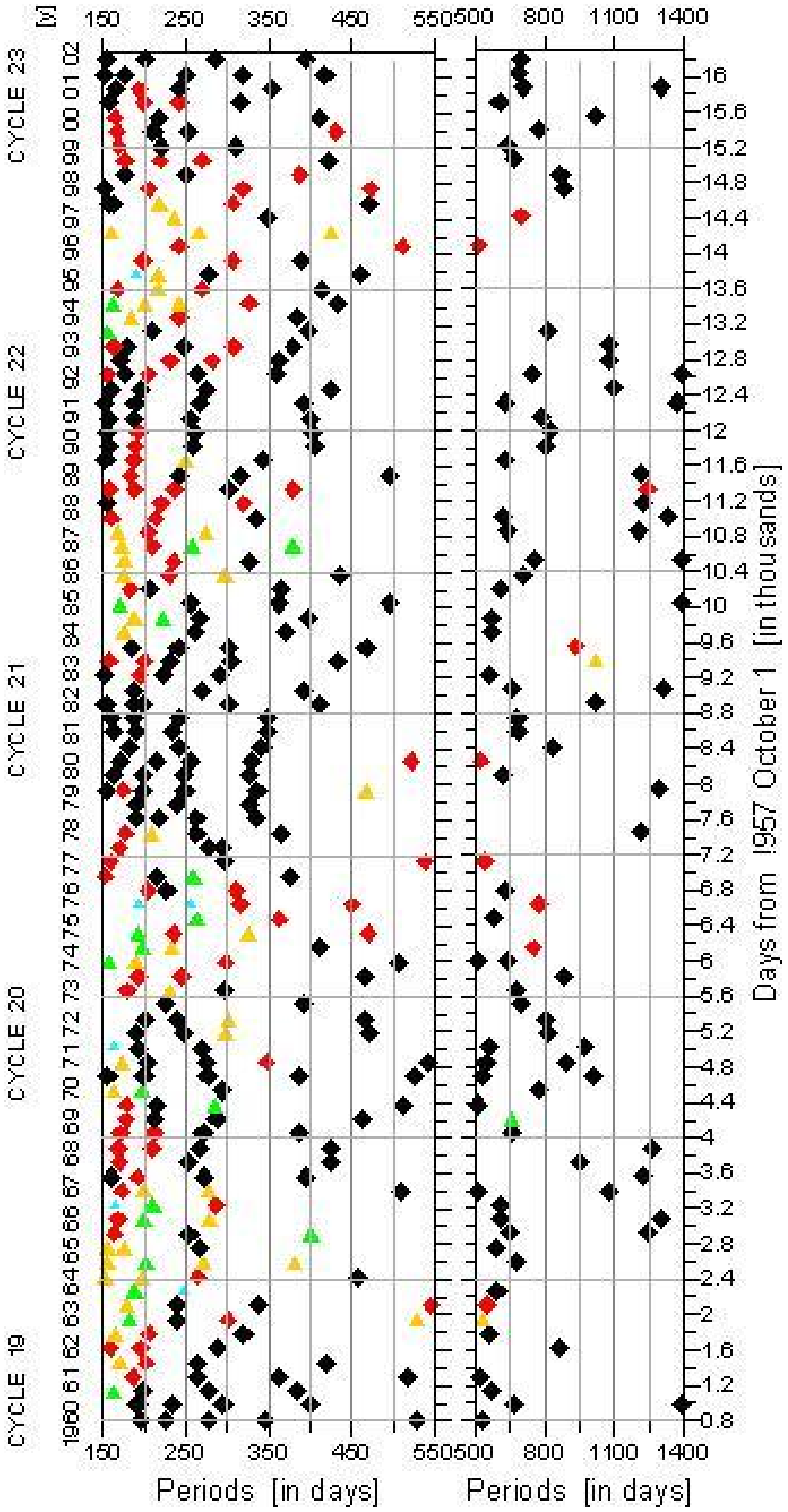

Fig. 7. —Amplitudes of harmonics with periods between 150 and 1400 days (near the period 500 days the scale of X axis is changed) found in 96 windows each having 1620 days width. The different colors are used to mark the amplitudes belonging to five equal number classes. The largest amplitudes are marked by black symbols and then for decreasing amplitudes red, yellow, green and blue colors are used. The Y-position of the symbol indicates the center of the window and the horizontal lines of grids are drawn in the distances equal to the window dimension.

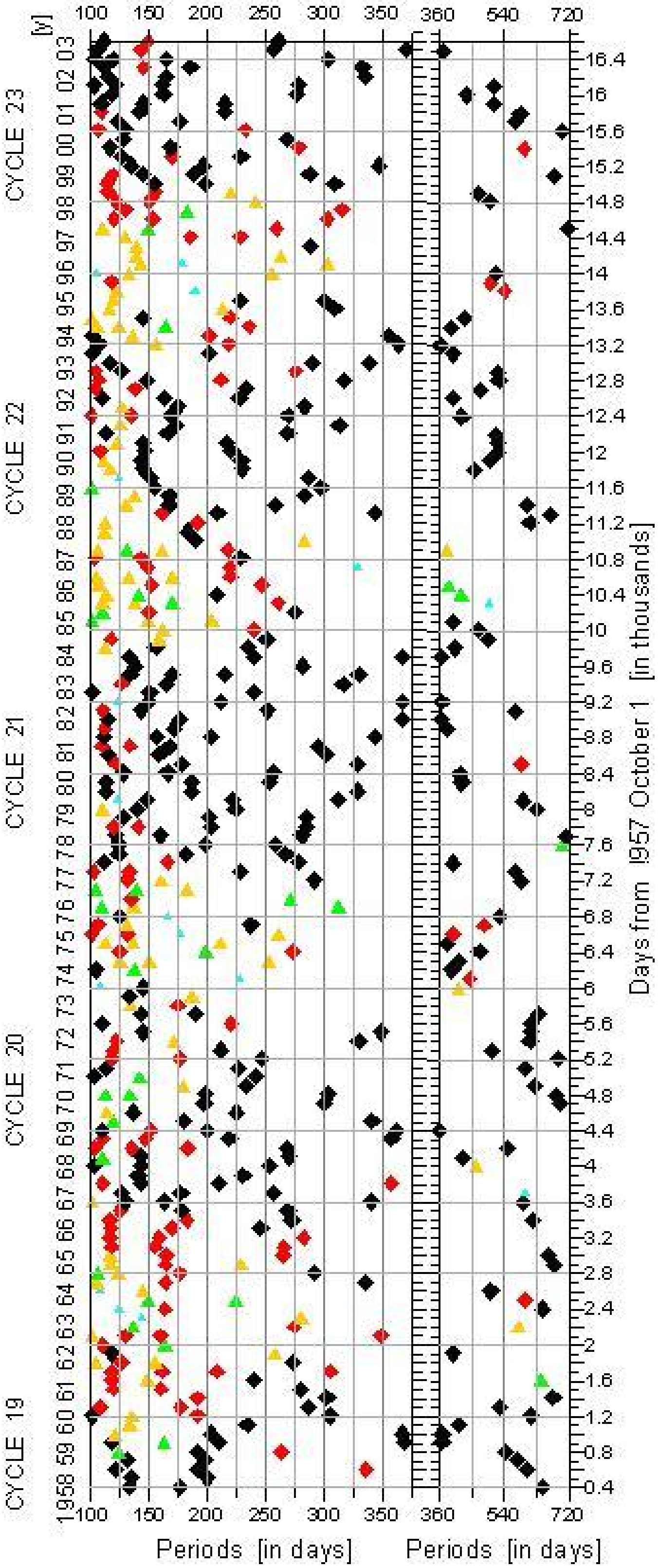

FIG. 8. —Amplitudes of harmonics with periods between 100 and 720 days (near the period 360 days the scale of X axis is changed) found in 163 windows each having 810 days width. The same scheme of colors as in Figure 6 is used to mark the amplitudes belonging to five equal number classes. The Y-position of the symbol indicates the center of the window and the horizontal lines of grids are drawn in the distances equal to the window dimension.

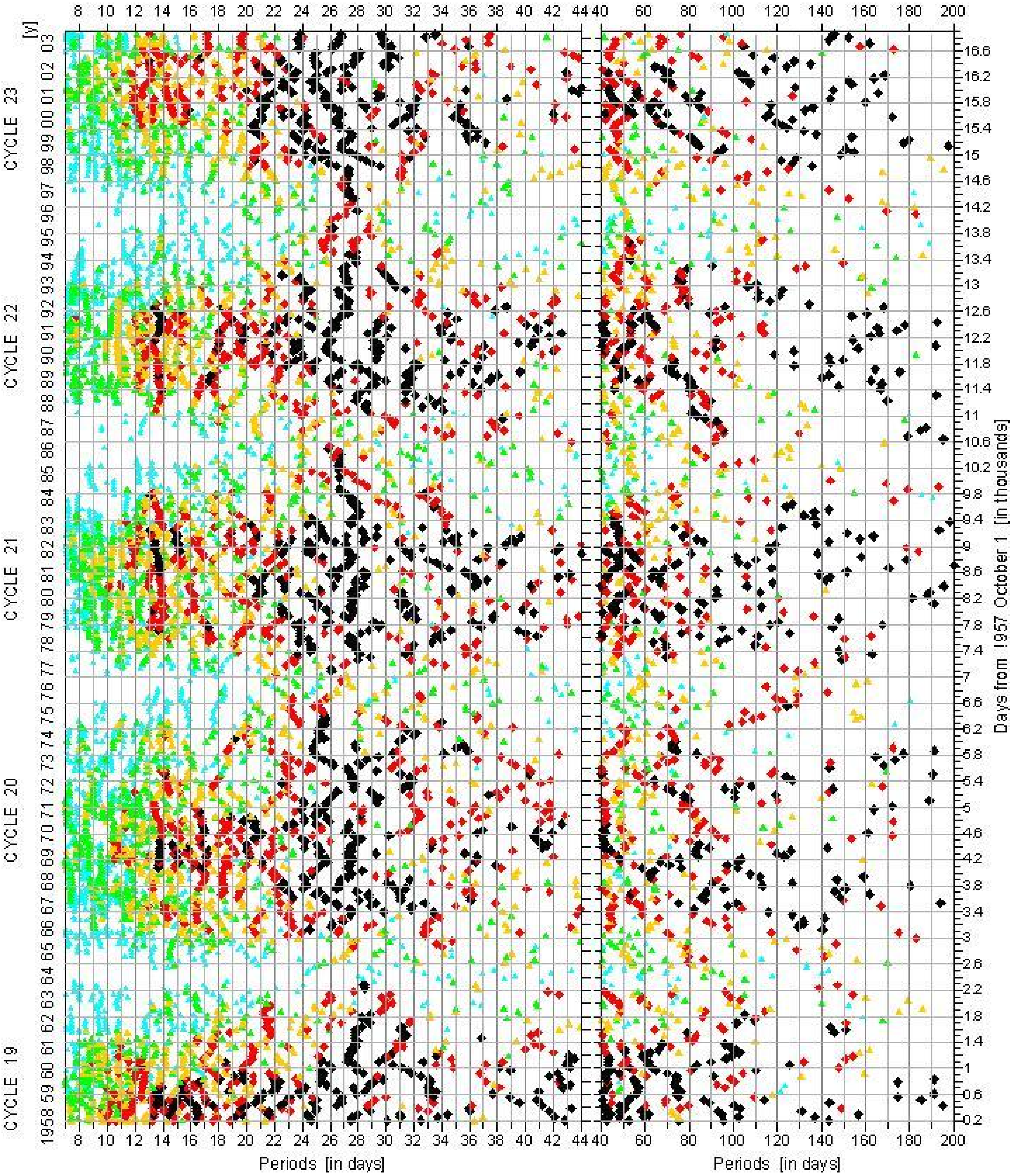

FIG. 9. —Amplitudes of harmonics with periods between 7 and 200 days (near the period 40 days the scale of X axis is changed) found in 371 windows each having 405 days width. The same scheme of colors as in Figure 6 is used to mark the amplitudes belonging to five equal number classes. The Y-position of the symbol indicates the center of the window and the horizontal lines of grids are drawn in the distances equal to the window dimension.

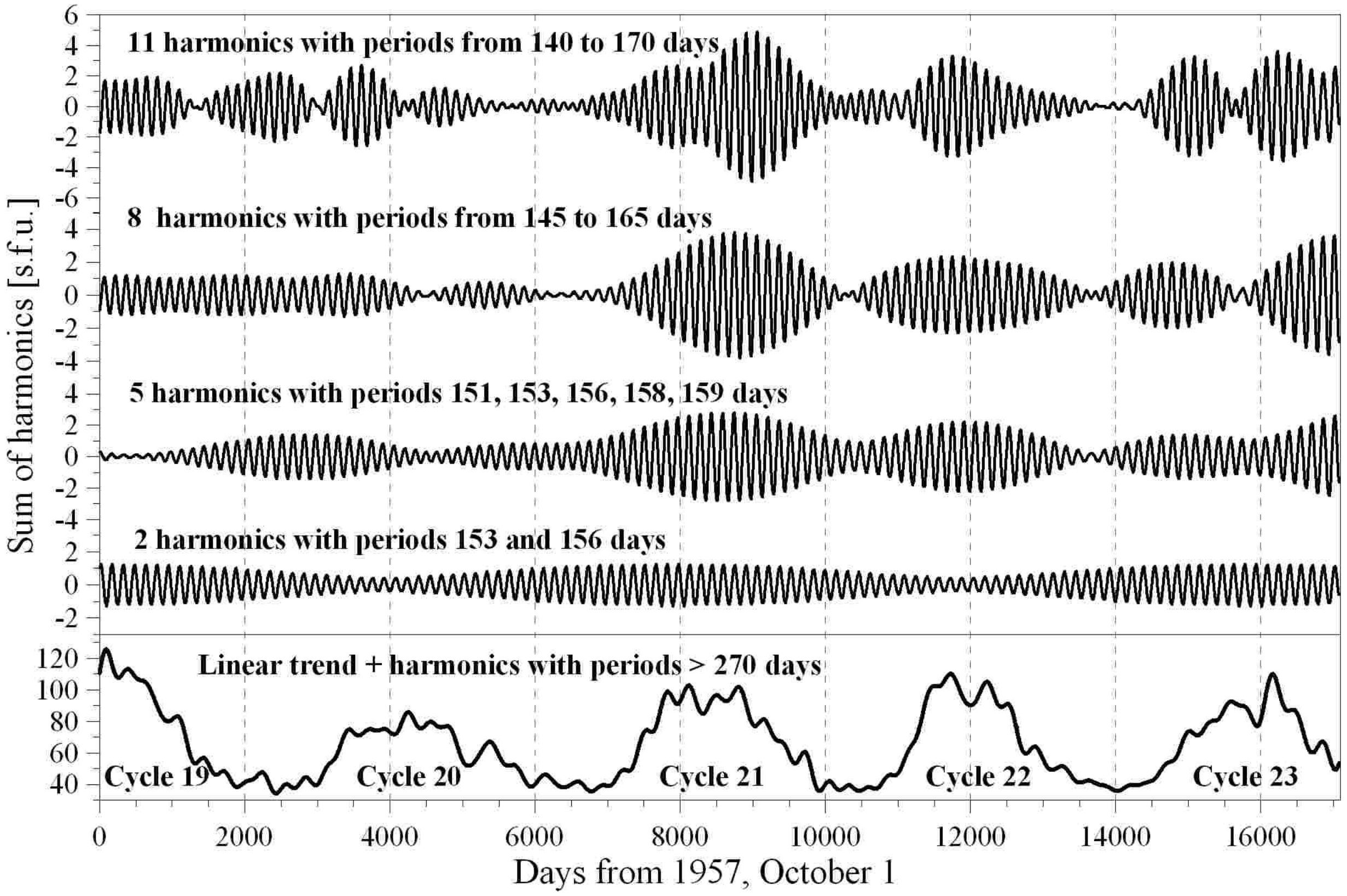

Fig. 10. —Sums of harmonics with various periods taken from the period interval 140 and 170 days.